# Effect of Annealing Temperature on Minimum Domain Size of Ferroelectric Hafnia

Seokjung Yun[#], Hoon Kim[#], Myungsoo Seo, Min-Ho Kang, Taeho Kim, Seongwoo Cho, Min Hyuk Park, Sanghun Jeon, Yang-Kyu Choi, and Seungbum Hong[*]

[a]*Department of Materials Science and Engineering, KAIST, Daejeon, 34141, Republic of Korea*

[b]*KAIST Institute for NanoCentury (KINC), KAIST, Daejeon, 34141, Republic of Korea*

[c]*Department of Materials Science and Engineering, KAIST, Daejeon, 34141, Republic of Korea*

[d]*Department of Materials Science and Engineering, Seoul National University, Seoul, Republic of Korea*

# Abstract

Here, we optimized the annealing temperature of HZO/TiN thin film heterostructure via multiscale analysis of remnant polarization, crystallographic phase, minimum ferroelectric domain size, and average grain size. We found that the remnant polarization was closely related to the relative amount of the orthorhombic phase whereas the minimum domain size was to the relative amount of the monoclinic phase. The minimum domain size was obtained at the annealing temperature of 500 ℃ while the optimum remnant polarization and capacitance at the annealing temperature of 600℃. We conclude that the minimum domain size is more important than the sheer magnitude of remnant polarization considering the retention and fatigue of switchable polarization in nanoscale ferroelectric devices. Our results are expected to contribute to the development of ultra-low-power logic transistors and next-generation non-volatile memory devices.

## Introduction

Hf$_{0.5}$Zr$_{0.5}$O$_2$ (HZO), which has ferroelectric properties through Zr doping with the conventional gate dielectric material HfO$_2$, has attracted much interest due to its potential application to next generation non-volatile memory devices and logic circuits. In addition, HZO is more suitable for the fabrication of FETs owing to its higher scalability, good thermal stability, and compatibility with Si processes than the previously introduced Pb(Zr,Ti)O$_3$ (PZT), SrBi$_2$Ta$_2$O$_9$ (SBT) and BaTiO$_3$ (BTO) ferroelectric materials.

Hafnia can have ferroelectric properties through various dopants such as Zr [10], Si [11], La [12], Y [13], Al [14], Gd [15], and Sr [16], as well as by post heat treatment and strain engineering between electrodes and oxide films. Through the process introduced above, hafnia with ferroelectric properties shows the formation of the non-centrosymmetric orthorhombic phase with a space group of *Pca*2$_1$, unlike conventional HfO$_2$ [17]. Because of this non-centrosymmetric orthorhombic phase, Park *et al*. reported ferroelectricity and non-ferroelectricity depending on the Zr concentration of HZO [18]. In addition, it was reported that the other dopants such as Y, Al, Gd, Sr, and La induce ferroelectric properties at a particular dopant concentration and heat treatment temperature [18]. The ferroelectric properties are lost due to the appearance of M-phase replacing O-phase above a specific temperature [19]. Furthermore, the strain between the electrode and the corresponding ferroelectric thin film affects the ferroelectric properties [20]. In the semiconductor process, as various temperatures, dopant concentrations, and electrodes can be used, it is crucial to find the optimum process condition to fabricate ferroelectric HZO thin with a high yield.

Many research groups have reported on the macroscopic characteristics of ferroelectrics, but only a few have analyzed the origin of ferroelectricity from the perspective of the interplay between domain configuration, defect distribution, polarization and screening charges [17,18,21]. To understand the polarization switching mechanism and retention or fatigue behavior of ferroelectric $HfO_2$ materials at the nanoscale, piezoresponse force microscopy (PFM) is one of the most powerful tools as it can visualize both out-of-plane and in-plane polarization vectors [22,23]. However, the challenge lies in the fact that the piezoelectric coefficient, which is the marker for polarization vector in PFM, is less than 10 pm/V. Stolichnov *et al*. reported $d_{eff}$ of HZO to be 5 pm/V, and Liam *et al*. reported effective piezoelectric coupling coefficient ($d_{eff}$) of < 2 pm/V in Si-doped hafnia [24,25]. The low piezoelectric coefficient results in low signal to noise ratio in PFM images, which makes the analysis difficult due to the various artifacts contributing to the PFM signal [23].

Here, we designed the voltage profile for PFM imaging of HZO thin films, which were fabricated with various rapid thermal annealing temperatures. We successfully analyzed the domain configuration of HZO thin films with different voltage pulse amplitude and duration to find the optimal process condition for minimizing the written domain size.

## Experiment

### Fabrication of $Hf_{0.5}Zr_{0.5}O_2$ thin films

8.5 nm thick $Hf_{0.5}Zr_{0.5}O_2$ (HZO) films were deposited on 200 nm thick TiN electrode as the bottom electrode/1.5 nm $SiO_2$ interfacial layer, using atomic layer deposition (ALD). Tetrakis(ethylmethylamido)hafnium(IV) (TEMAHf), Tetrakis(ethylmethylamido)zirconium(IV) (TEMAZr), and $O_3$ were used as Hf precursor, Zr precursor, and oxygen reactant, respectively. A 200 nm TiN electrode was deposited on top of the device via physical vapor deposition (PVD). The top electrode of TiN was wet etched using an $NH_4OH:H_2O_2:H_2O$ solution at a concentration ratio of 1:2:5 and the pattern size was 30~500 $\mu m$. The ferroelectric HZO film was crystallized by post-metal rapid thermal annealing (RTA) in the temperature range from 400 to 900℃.

We confirmed the thickness of 9.8 nm thick HZO through a cross-sectional TEM image, which was deposited by ALD on the TiN bottom electrode followed by the deposition of the TiN top electrode. The atomic percent ratio of the HZO film was confirmed by X-ray photoelectron spectroscopy (XPS) (Sigma Probe, Thermo VG Scientific) depth profiling, which was around 1:1 in terms of Hf: Zr ratio. The crystal structure of the HZO films was analyzed using grazing incidence X-ray diffraction (GIXRD) (Rigaku, SmartLab) after the post-metal annealing. The polarization-electric field (P-E) hysteresis loops and the capacitance-voltage (C-V) characteristics were obtained by Precision LC II.

### Piezoresponse force microscopy (PFM) imaging of $Hf_{0.5}Zr_{0.5}O_2$ thin films

To understand the nano-scale polarization and domain properties of HZO, PFM imaging was conducted on the HZO surface after removing the top electrode. We used NX-10 (Parks system) and Cypher ES (Asylum Research, Oxford Instruments) to measure the grain size and conduct piezoresponse force microscopy (PFM) analysis, respectively. To measure the grain size of HZO thin films with RTA temperature between 500 and 800 ℃, we conducted non-contact mode AFM imaging using an Al-coated AFM tip (AC160TS, Olympus). Scan size and speed were 500 nm × 500 nm and 0.5 Hz, respectively. In the case of PFM analysis, we conducted dual ac resonance tracking (DART) PFM using a Pt/Ir-coated AFM tip (PPP-EFM, Nanoworld) [26,27].

The ac modulation voltage and loading force were 1V and 40 nN, respectively. The drive frequencies were near the contact resonance between 320 and 350 kHz. To measure the domain size, we conducted the nanoscale bit formation process. First, we performed background poling as the first step, where we applied −7 V and +7 V respectively for up and down switching over area of 3 μm × 3 μm. For dot poling, we applied +7 V and +8 V on the up-domain region and −7 V and −8 V on the down-domain region as a function of pulse time, respectively. After the poling process, we scanned and confirmed that the nanoscale bits were successfully formed. The imaging experiments were repeated more than 10 times.

## Results and Discussion

### Macro-scale characteristic of HZO thin films with various RTA temperatures

The thickness of the HZO thin film is 98 Å, as confirmed by transmission electron microscopy (TEM) in Fig. 1 (a). Fig. 1 (b) and (c) show the grazing incidence X-ray diffraction (GI-XRD) spectrum and the relative ratio of monoclinic, orthorhombic, and tetragonal phases, respectively. The crystalline peaks started to appear from 500 ℃ up to 900 ℃. Park *et al*. reported that the 2 theta peaks between 28 and 31º could be assigned as m(-111), o(111), t(011), and m(111), respectively [26]. As such, one can deconvolute the overlapped peaks into m(-111), o(111), t(011), and m(111) that are located at 28.54º, 30.4º, 30.8º, and 31.64º [28]. The relative ratio of each phase obtained by Gaussian peaks from each RTA temperature is shown in Fig. 1 (c). The m-phase increases with increasing RTA temperature, the o-phase increases up to 700 °C and then slightly decreases at 800 °C. Our results are in good agreement with the prior XRD results [28]. We conclude that the optimum temperature range for o-phase crystallization is between 600 and 700 °C.

In ferroelectric HZO MFM structure, the electrode is one of the critical factors for high ferroelectricity and polarization switching behavior. Among the various electrodes, TiN has been reported as one of the most suitable electrodes to have high ferroelectricity in HZO capacitors.

P-V and C-V characteristics were measured using Precision LC II equipment. Among the samples fabricated by various RTA temperatures, the well-defined ferroelectric properties with a large $2P_r$ of >30 $\mu$C/cm$^2$ were observed in the case of the HZO films with RTA at 600°C, as shown Fig. 2 (a), from P-E hysteresis loops. In the case of RTA at 900°C, the capacitor failed to display a well-defined hysteresis loop and showed a doughnut shape, a signature of a leaky ferroelectric film.

Figure 2 (b) shows the butterfly-like feature of the capacitance-voltage (C-V) characteristic of the HZO thin films at various RTA temperatures. P-E hysteresis and C-V measurements of the HZO samples with RTA between 500 and 800°C, show well-defined ferroelectric characteristics, of which results are in good agreement with prior report [29].

Figure 2 (c) shows the wake-up effect, which is a typical characteristic of HZO thin films where $P_r$ starts to increase from $10^2$ cycles to $10^4$ cycles of polarization switching. Based on macroscale characterization via XRD, P-V, and C-V analysis, we found that the optimum RTA temperature was 600°C.

**Nanoscale characterization of HZO thin films as a function of RTA temperature**

We conducted AFM experiment to complement the optimization of annealing temperature, which was based on the macroscale analysis. In the case of HZO thin films, unlike conventional ferroelectric materials such as $PbTiO_3$ (PTO), PZT, BTO, and P(VDF-TrFE) (Poly(vinylidene fluoride-trifluoroethylene), the polarization poling process of ferroelectric domain is challenging [30-34].

The reasons behind the challenge of poling process are as follows. First, the piezoelectric coefficient of HZO thin film is ~10 pm/V [35], which is very low compared to that of the conventional piezoelectric material such as PZT, PTO, and $BiFeO_3$ (BFO) [36-38]. This means that PFM signal may be influenced by other artifacts such as electrostatic force induced vibration [23]. Furthermore, HZO thin film is likely to have defects such as non-ferroelectric layer, oxygen vacancy, and electron trap inside the film during deposition and heat treatment [39]. Since HZO thin films have low $d_{33}$ compared to other ferroelectric materials with potential imprint in the hysteresis loop, the PFM amplitude and phase of HZO thin films may suffer from low signal-to-noise ratio [40]. To remove the internal bias between the electrode and HZO thin films, we applied an additional bias (1~2 V) to the tip to neutralize the internal bias. After removing the internal bias, we could observe clear PFM amplitude and phase signals with domain boundaries as shown in Fig. S1.

We measured the domain size using the method introduced by Woo *et al.* [41]. Fig. 3(a) shows the sequences of nanoscale bit formation. We used lithography function where we could form up- and down-domains by applying predefined bias over a region of 3 μm × 3 μm as shown in Fig. 3(a). More specifically, a negative bias was applied to the AFM tip to form up-domains over a rectangle region of 1.5 μm × 3 μm, and a positive bias was applied to form down-domains over a rectangle region of 1.5 μm × 3 μm next to the up-domain region.

Subsequently, we formed a matrix of dot array by changing the amplitude and duration of voltage pulse applied to the AFM tip. After the matrix formation, we acquired the topography (left), PFM amplitude (middle), and phase (right) images as shown in Fig. 3(b). The upward (downward) polarizations show bright (dark) contrast without inducing any change in the topography. The matrix formation was repeated on the HZO thin films with RTA conducted between 500 and 800 ℃.

Fig. 4 shows the domain dot size (i.e. domain diameter) of RTA 600℃ sample as a function of pulse time (500 μs, 5 ms, 50 ms, 500 ms, and 1 s) for different bias voltages of +7, +8, −7 and −8 V, which was measured from PFM phase and amplitude images by taking the full-width at half maximum (FWHM) value of each line profile [41]. The reason why the coercive voltage is much larger in PFM hysteresis loops than that in P-E hysteresis loops can be explained by the inhomogeneous electric field generated by the PFM tip and critical size of domain nucleus [41]. The lateral domain size from amplitude and phase images was linearly proportional to logarithmic value of the pulse width.

Fig. 5(a)-(d) shows the grain size of HZO thin films as a function of RTA temperature. We found that the grain size increases as a function of RTA temperature, as shown in Fig. 5 (e). The orthorhombic phase (O-phase) is theoretically unstable, which is responsible for the unexpected ferroelectric property of HZO thin films. As such, grain size and surface energy are important factors in understanding the appearance of the O-phase in HZO thin films [42,43]. As the grain size increases, the portion of M-phase increases while the portion of O-phase decreases as shown in Fig. 1. We speculate that the low surface energy of M-phase played an important role in increasing the grain size at the expense of the O-phase as the RTA temperature

increased.

Fig. 6(a) shows the correlation between the minimum domain size and the RTA temperature. We observed that the minimum domain size increases with RTA temperature. If the M-phase is mostly distributed at the surface, then the surface layer can act as a dead layer, which will decrease the effective electric field applied over the O-phase. At the same time, it will act like a lens through which the electric field will distribute more sparsely over the O-phase layer. This hypothesis can explain the correlation between the minimum domain size, grain size and the RTA temperature.

We found that the minimum domain size was 36.6 nm at RTA temperature of 500℃ as shown in Fig. 6 (a). This is the smallest domain of HZO observed by PFM reported up to date. The domain size increased by 39.4 nm when the RTA temperature increased by 100℃. This information helps the engineers to design the RTA process for ferroelectric hafnia memory devices, as it will influence the processing window and the resulting yield of the device.

Fig. 6 (b) shows the correlation between domain size, $P_r$, M-phase, and RTA temperature. In the case of $P_r$ and M-phase, the optimal condition lies between 600 and 700℃, whereas, in the case of the minimum domain size, the optimum condition can be found below 600 ℃ as described in Table 1. Kim *et al.* reported that the memory window and reliability of the device deteriorated when $P_s$ increased in the Fe-FETs using ferroelectric hafnia [44]. Based on our results and Kim *et al*.'s work, we conclude that the RTA temperature of 600 ℃ or lower is more suitable for high-performance, high-endurance, and highly integrated Fe-FETs using ferroelectric hafnia in the future.

**Conclusions**

We investigated the minimum domain size of ferroelectric hafnia thin films as a function of RTA temperature using dual ac resonance tracking (DART) PFM. We found that the minimum domain size scales linearly with the RTA temperature. The maximum capacitance and remnant polarization were found at 600 and 700 ℃, whereas the minimum domain size was found at 500 ℃. The above findings were analyzed in terms of the ratio of monoclinic phase to orthorhombic phase and its role in domain size and grain size. Our findings provide direct evidence of high-density ferroelectric memory devices based on nanoscale domain switching.

**Acknowledgement**

This work was supported by the Samsung Electronics Semiconductor Research Project funded by Samsung Electronics, and the National Research Foundation of Korea (NRF) grant funded by the Korea government (MSIT) (No. 2020R1A2C201207811).

**(a)**

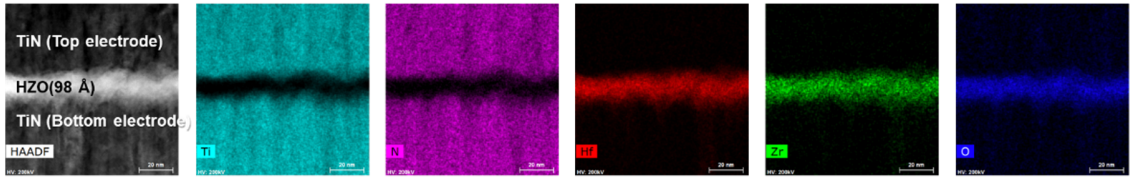

**(b)**

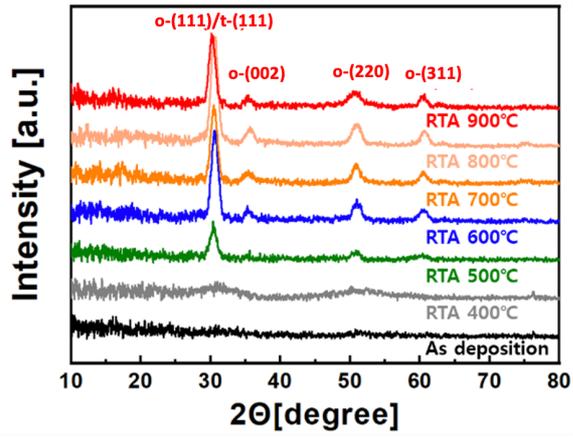

**(c)**

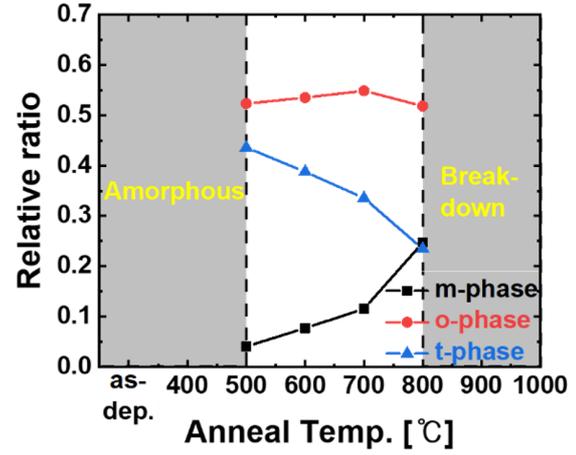

**Figure 1.** (a) transmission electron microscopy (TEM) images of HfZrO M-F-M structures, (b) GI-XRD, (c) relative ratio of HZO thin films with various RTA temperature

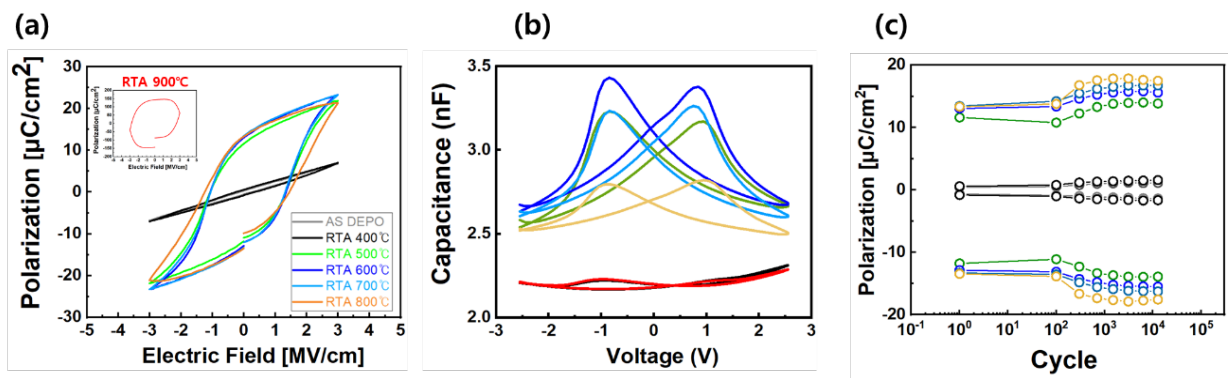

**Figure 2.** (a) Polarization–Electric field (P-E) curve, (b) Capacitance–Voltage (C-V) curve, and (c) cycle properties of HZO films with various RTA temperature.

**(a)**

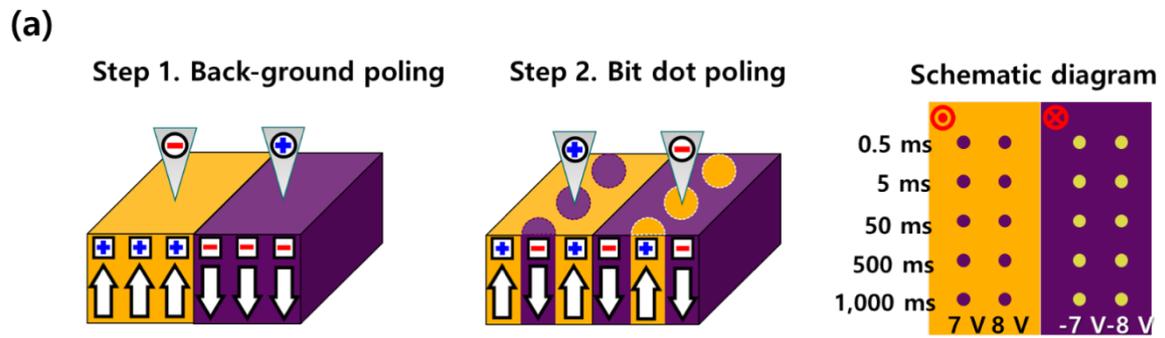

**(b)**

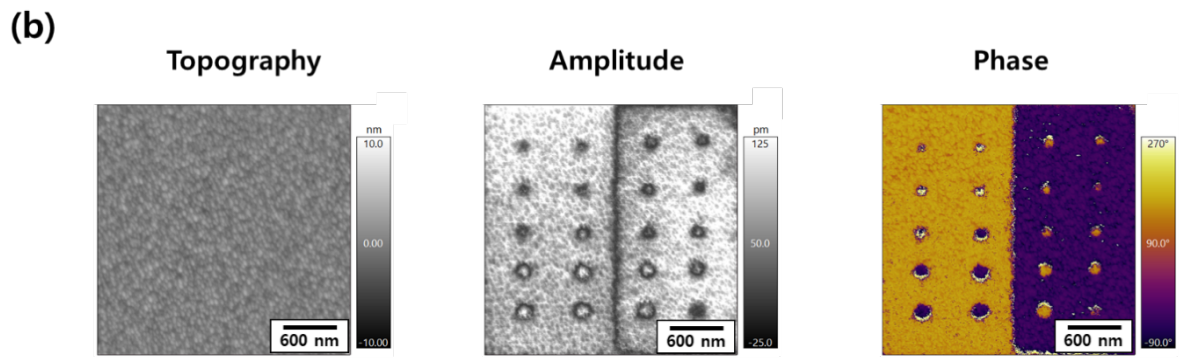

**Figure 3.** (a) Process sequences of nano-scale bit formation, (b) After poling process, the topography(left), amplitude(middle), and phase(right) images

**(a)**

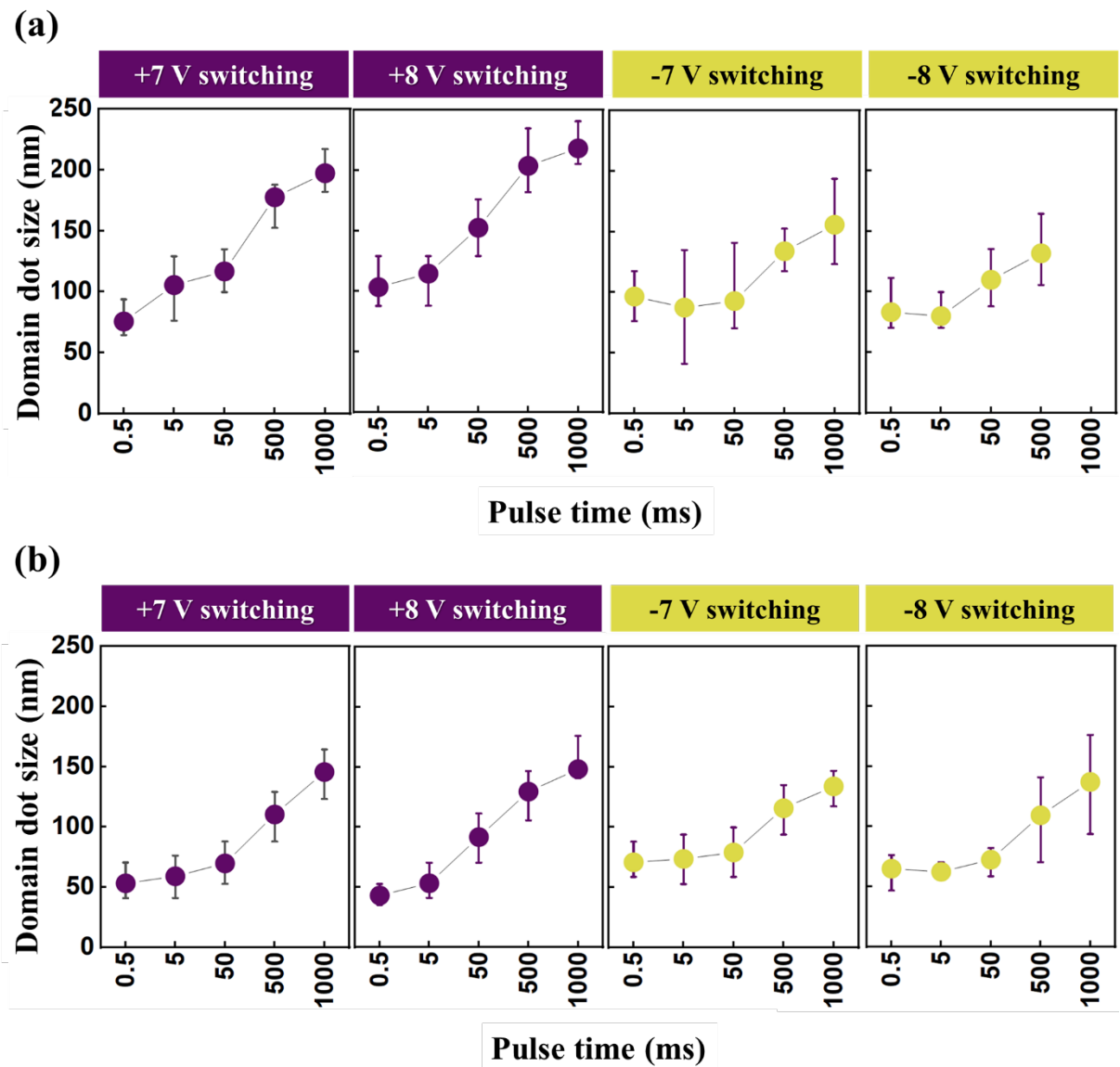

**(b)**

**Figure 4.** PFM (a) phase, (b) FHWM domain size from phase image, (c) amplitude, and (d) FHWM domain size from amplitude image of HZO at RTA 600 ℃,

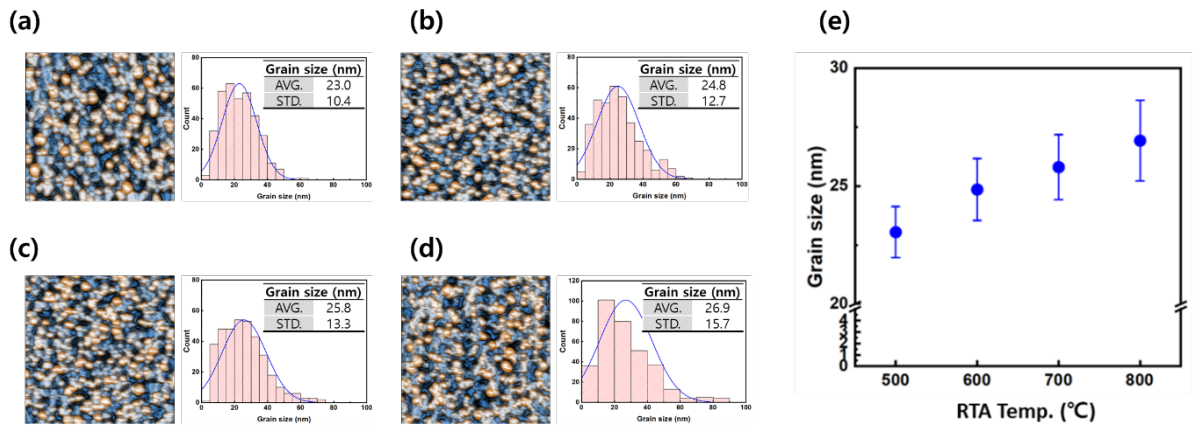

**Figure 5.** Grain size of HZO thin film from (a) RTA 500 ℃, (b) RTA 600 ℃, (c) RTA 700 ℃, (d) RTA 800 ℃ temperatures, and (e) correlation between domain size and RTA temperatures

**(a)**

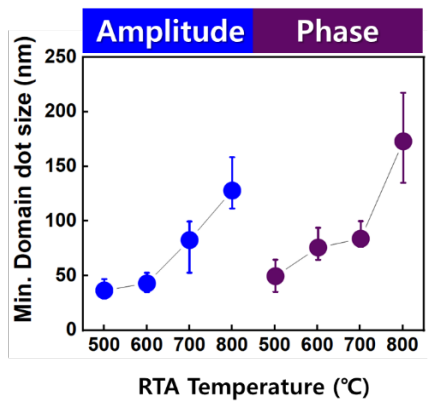

**(b)**

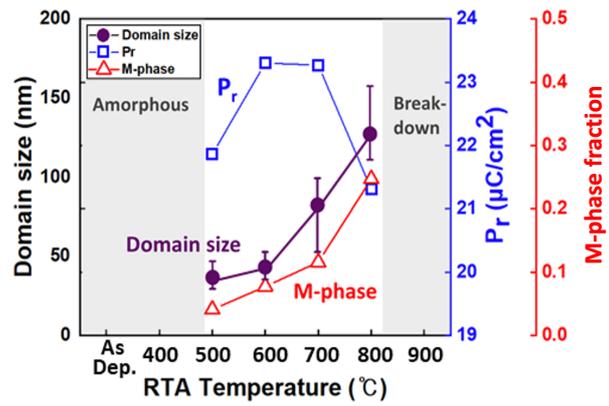

**Figure 6.** (a) Correlation between domain size and RTA temperatures and (b) correlation between domain size, $P_r$, M-phase, and RTA temperatures.

**Table 1.** Summary of O-phase, $P_r$, Min. grain size, and domain size from experimental results.

| | RTA temperature | | | |
|---|---|---|---|---|
| | 500 ℃ | 600 ℃ | 700 ℃ | 800 ℃ |
| **O-phase (%)** | 52 | 54 | 55 | 52 |
| **M-phase (%)** | 4 | 8 | 12 | 25 |
| **Remnant polarization (μC/cm²)** | 21.9 | 23.3 | 23.3 | 21.3 |
| **Average grain size (nm)** | 23.0 | 24.8 | 25.8 | 26.9 |
| **Minimum domain size (nm)** | 36.6 | 43.2 | 87.5 | 120.6 |

Supporting Information for

# Effect of Annealing Temperature on Minimum Domain Size of Ferroelectric Hafnia

Seokjung Yun[#], Hoon Kim[#], Myungsoo Seo, Min-Ho Kang, Taeho Kim, Seongwoo Cho,
Min Hyuk Park, Sanghun Jeon, Yang-Kyu Choi, and Seungbum Hong[*]

[a]*Department of Materials Science and Engineering, KAIST, Daejeon, 34141, Republic of Korea*

[b]*KAIST Institute for NanoCentury (KINC), KAIST, Daejeon, 34141, Republic of Korea*

[c]*Department of Materials Science and Engineering, KAIST, Daejeon, 34141, Republic of Korea*

[d]*Department of Materials Science and Engineering, Seoul National University, Seoul, Republic of Korea*

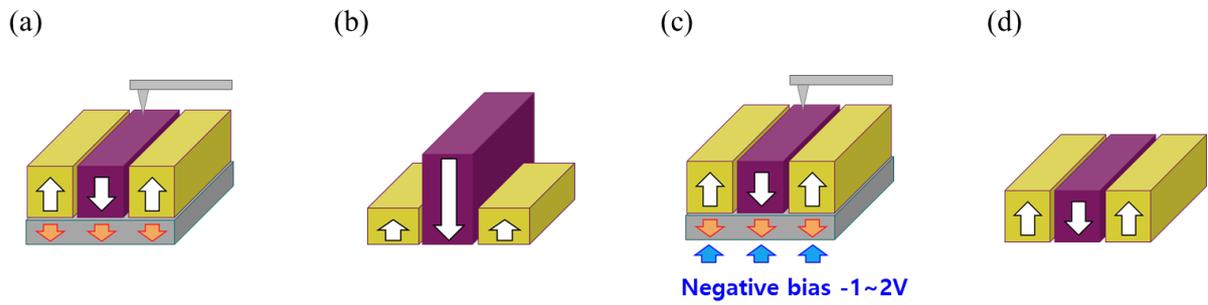

**Figure S1**. The schematic of (a) the domain from HZO thin film with internal bias, (b) the domain signal distorted due to internal bias, (c) the domain from HZO thin film applied bias with removed internal bias, and (c) the domain signal after removed internal bias

(a)                              (b)                              (c)

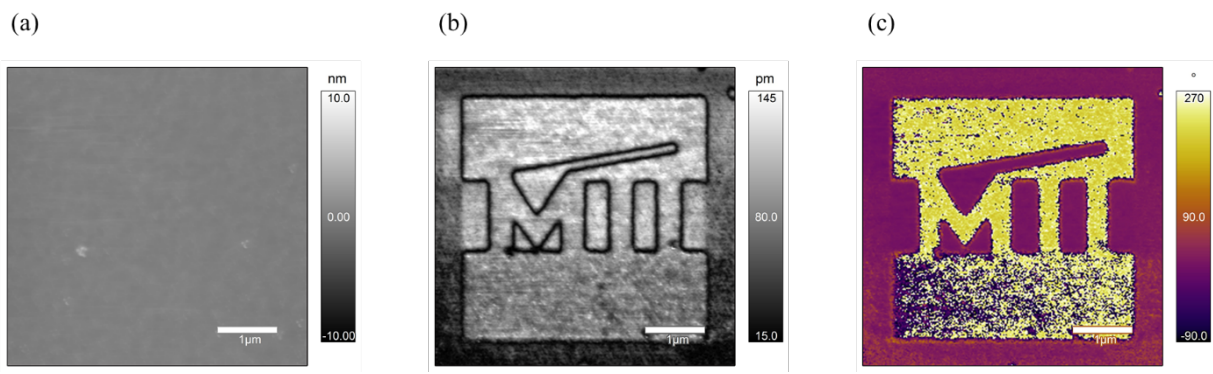

**Figure S2**. (a) topography, (b) amplitude, and (c) phase images from the HZO thin film after poling process with removed internal bias

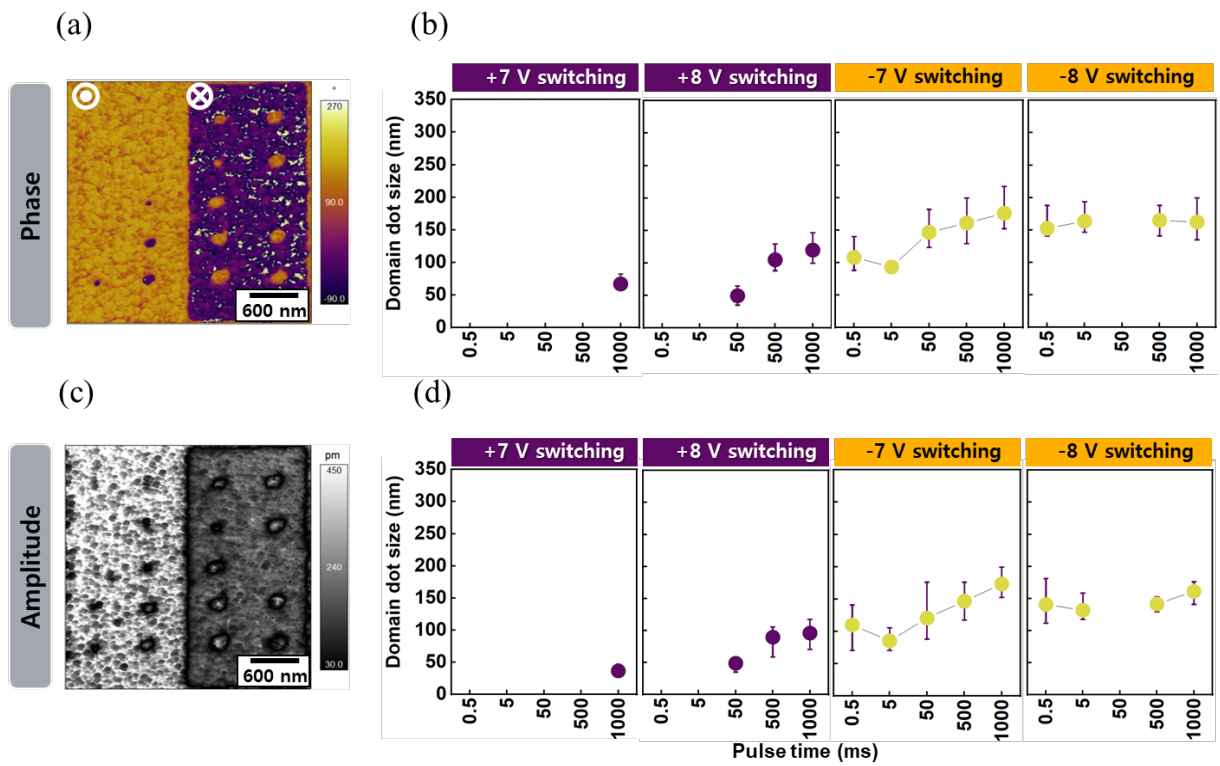

**Figure S3**. (a) phase image, (b) domain size of phase image, (c) amplitude image, and (d) domain size of amplitude images from the HZO thin film that annealed at 500 °C.

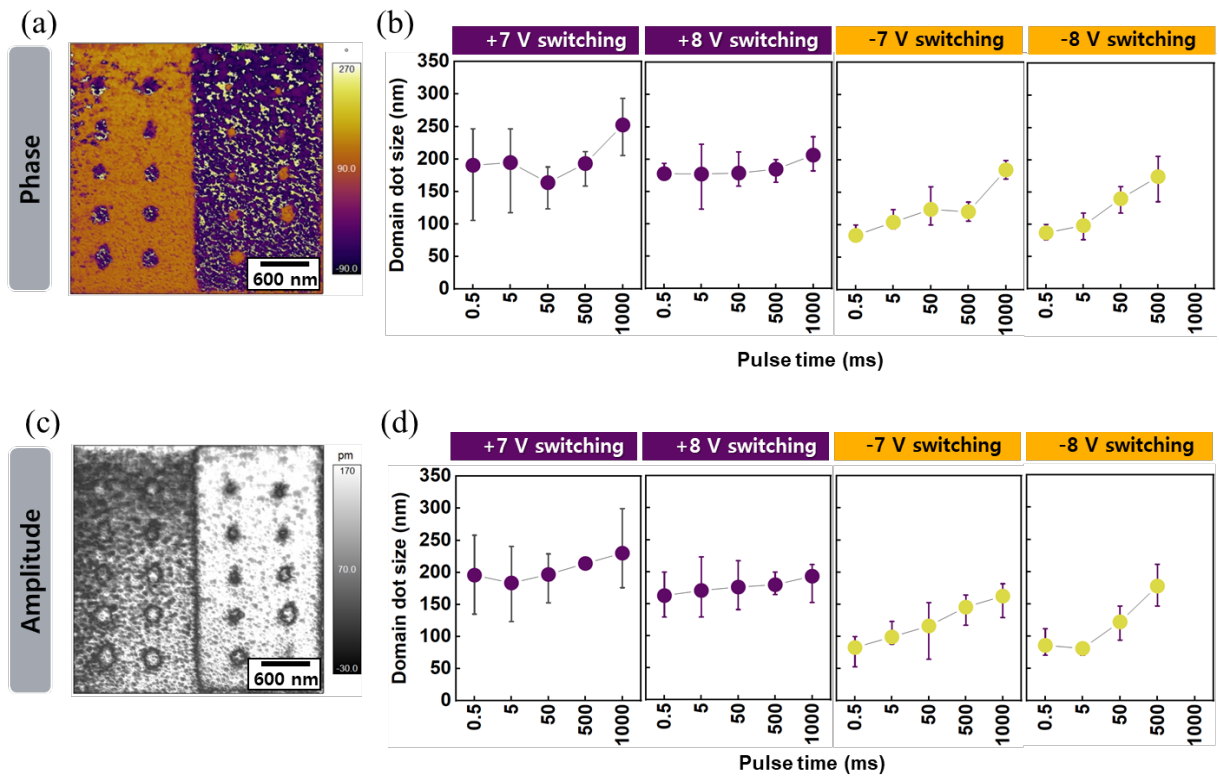

**Figure S4**. (a) phase image, (b) domain size of phase image, (c) amplitude image, and (d) domain size of amplitude images from the HZO thin film that annealed at 700 °C.

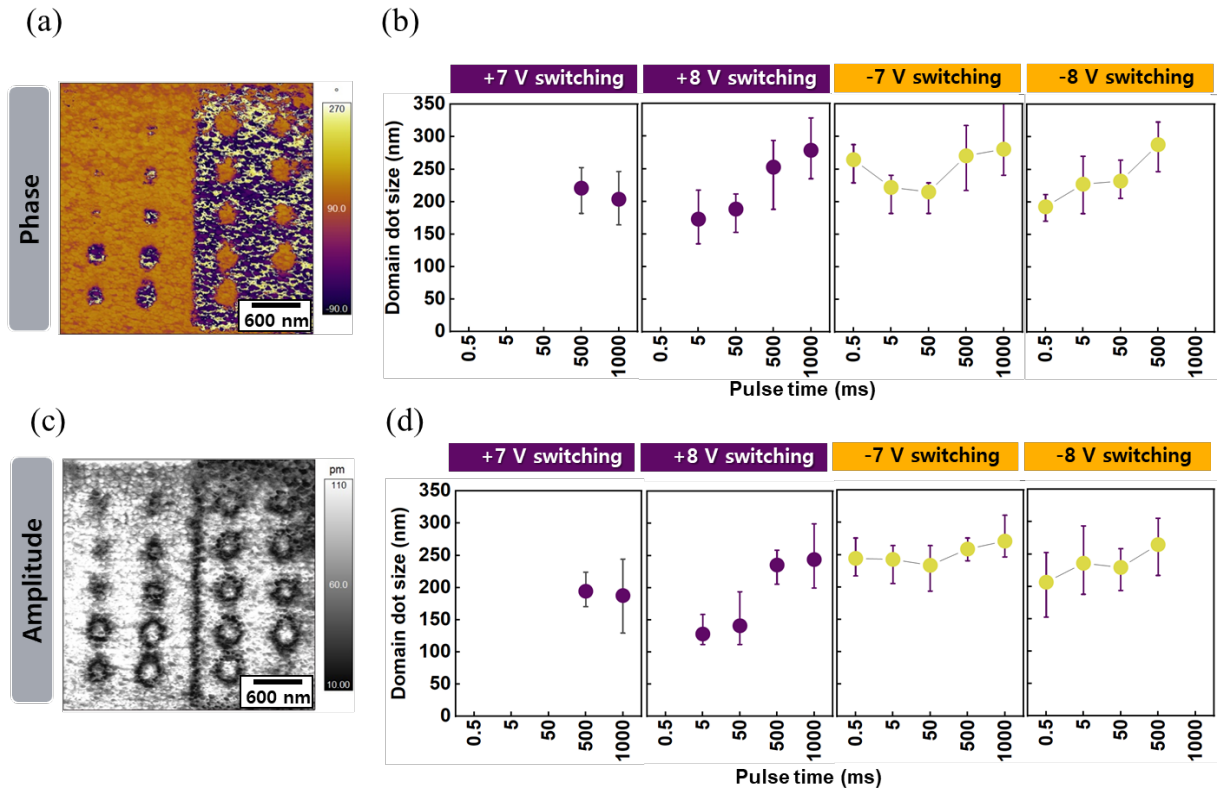

**Figure S5**. (a) phase image, (b) domain size of phase image, (c) amplitude image, and (d) domain size of amplitude images from the HZO thin film that annealed at 800 °C.

**Graphical Summary**

°

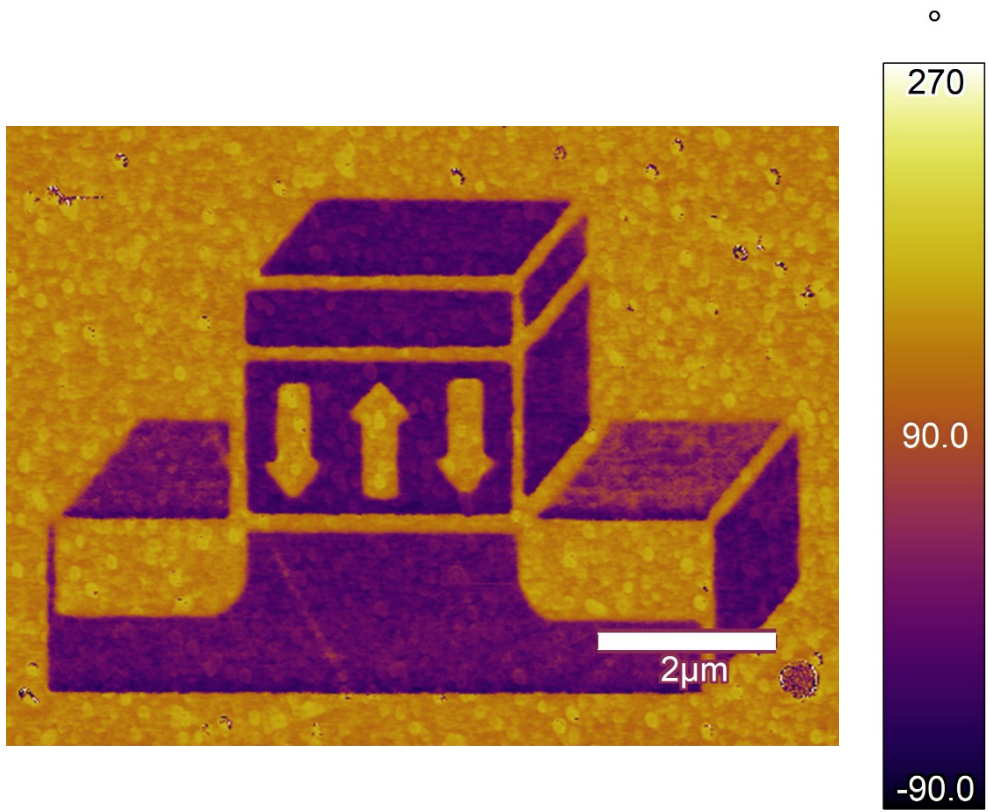

270

90.0

-90.0

2µm

pm

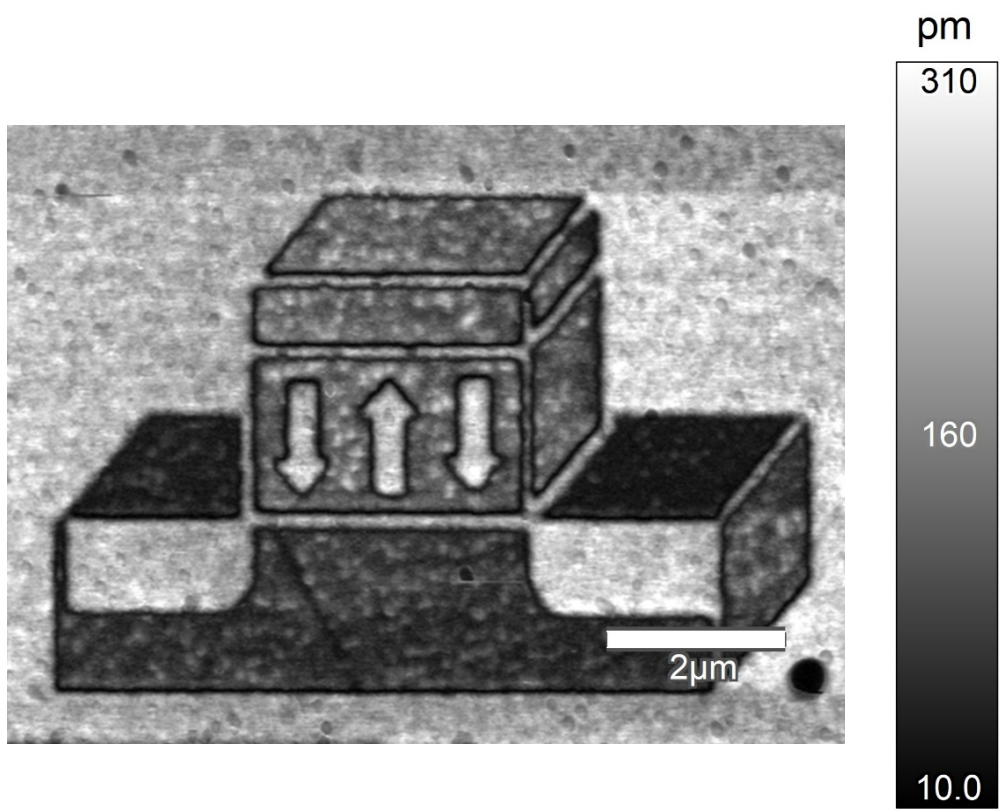

310

160

10.0

2µm